\documentclass[english]{article}
\usepackage[T1]{fontenc}
\usepackage[koi8-r]{inputenc}
\usepackage{amsmath}
\usepackage{amssymb}
\usepackage{esint}

\makeatletter
\@ifundefined{date}{}{\date{}}
\newtheorem{theorem}{Theorem}

\newtheorem{corollary}{Corollary}

\newtheorem{lemma}{Lemma}

\newtheorem{remark}{Remark}

\usepackage{babel}

\usepackage{babel}

\usepackage{babel}

\usepackage{babel}

\makeatother

\usepackage{babel}
\begin{document}

\title{Convergence to Gibbs equilibrium - unveiling the mystery}

\author{A. A. Lykov, V. A. Malyshev %
\thanks{Faculty of Mechanics and Mathematics, Lomonosov Moscow State University.
Vorobievy Gory, Main Building, 119991, Moscow Russia, malyshev2@yahoo.com%
}}
\maketitle
\begin{abstract}
We consider general hamiltonian systems with quadratic interaction
potential and $N<\infty$ degrees of freedom, only $m$ of which have
contact with external world, that is subjected to damping and random
stationary external forces. We show that, as $t\to\infty$, already
for $m=1$, the unique limiting distribution exists for almost all
interactions. Moreover, it is Gibbs if the external force is the white
noise, but typically not Gibbs for gaussian processes with smooth
trajectories. This conclusion survives also in the thermodynamic limit
$N\to\infty$. 
\end{abstract}

\section{Introduction}

One of the most important, hard and long-standing problems in non-equilibrium
classical statistical physics is the convergence to Gibbs equilibrium.
One can say even that the mathematical status of this problem had
always been a bit mysterious. No mathematical argument, for many-particle
systems, appeared to justify convergence to equilibrium for closed
deterministic systems. On the contrary, there were many examples (linear
systems, completely integrable systems and their non-linear perturbations
within KAM theory) showing exactly the contrary. Moreover, for finite
quantum systems with unitary dynamics it is obvious that there cannot
be any convergence to equilibrium - contact with external world is
absolutely necessary.

In despite of this, since Boltzmann and Gibbs, it has often been believed
that non-linear effects (particle collisions), inside the closed system,
could provide this convergence. Closed linear hamiltonian systems
were always considered as annoying, thus rare and uninteresting exception,
where the abundance of invariant subspaces and invariant tori prevents
the dynamic emergence of limiting Gibbs states.

Sometimes, this difficulty has been overcome by artificially introducing
specially chosen stochastic internal dynamics throughout all closed
system. Then sometimes it became possible to prove convergence to
Gibbs invariant measure.

We cannot and even do not intend to disprove such common belief. Our
goal is much more modest - we want to show that there can be an alternative
approach to the convergence problem. Namely, let us assume that completely
closed system is an idealization, and there is always some, even the
smallest possible, contact with the external media. Then, as we show
here for general systems with quadratic interaction, the situation
changes drastically - invariant subspaces and tori become dynamically
intermixed - and linear systems become not an exception, but a legal
member of the model community. <<Very small>> means for us that,
for example, only one (of $N$) degree of freedom contacts external
world. There were a series of papers by J. Lebowitz and colleagues
(see for example \cite{leb_rieder,leb_spohn,leb_bonetto_lukka} and
references therein), devoted to non-equilibrium models of one-dimensional
crystals with different assumptions and different goal.

We consider general linear hamiltonian system with $N$ degrees of
freedom and assume that one (or more) fixed degree of freedom is subjected
to damping and random stationary external force. We prove that if
the external force is the white noise then there is convergence to
Gibbs state. However, if the external force is a stationary random
process with smooth trajectories then <<typically>> it converges
to equilibrium but this equilibrium will not be Gibbs. This brings
the conclusion that the absence of memory in the external force may
be crucial for the convergence to Gibbs equilibrium. More interesting
(and more difficult to prove) is that this assertion holds also in
the thermodynamic limit - that is for the degrees of freedom far away
from the contacts with external world. <<Typically>> means generic
situation in the common sense and is accurately explained in the text.

Our paper puts also another question - why the closed deterministic
systems feels even the smallest influence from the boundary so sharply.
We think that the same should hold also for non-linear systems - collisions
can only accelerate or relax this influence. However we cannot prove
it now. Such sharp feeling of the boundary is rarely possible for
stochastic dynamics. Possibly this is the reason why the fundamental
physical laws are deterministic, not stochastic. We do not claim that
our scheme for the convergence is the only possible but we do not
know other possibilities.

\section{Necessary definitions}

We consider the phase space 
\[
L=L_{2N}=\mathbb{R}^{2N}=\{\psi=(\begin{array}{c}
q\\
p
\end{array}),q=(q_{1},...,q_{N})^{T},p=(p_{1},...,p_{N})^{T}\in\mathbb{R}^{N}\},
\]
($T$ denotes transposition, thus $q,p,\psi$ are the column vectors)
with the scalar product 
\[
(\psi,\psi')_{2}=\sum_{i=1}^{N}(q_{i}q_{i}'+p_{i}p_{i}')
\]
It can be presented as the direct sum 
\begin{equation}
L=l_{N}^{(q)}\oplus l_{N}^{(p)}\label{direct_sum}
\end{equation}
of orthogonal coordinate and momentum subspaces, with induced scalar
products $(q,q')_{2}$ and $(p,p')_{2}$ correspondingly. We distinguish
several degrees of freedom , say 
\[
\Lambda^{(m)}=\Lambda^{(N,m)}=\{N-m+1,...,N\}\subset\Lambda=\{1,...,N\},1\leq m\leq N,
\]
(we shall call the set $\Lambda^{(m)}$ the boundary of $\Lambda$)
and consider the dynamics defined by the system of $2N$ stochastic
differential equations 
\begin{equation}
\frac{dq_{k}}{dt}=p_{k}\label{system_2N}
\end{equation}
\[
\frac{dp_{k}}{dt}=-\sum_{l=1}^{N}V(k,l)q_{l}-\alpha\delta_{k}^{(N,m)}p_{k}+F_{t,N+k}
\]
where $k=1,...,N$, $V=(V(k,l))$ is a positive definite $(N\times N)$-matrix,
$\delta_{k}^{(N,m)}=1$ if $k>N-m$ and zero otherwise. It is convenient
to define the $2N$-vector $F_{t}$ with the components: $F_{t,k}=0,k\leq2N-m$,
and $F_{t,k},k>2N-m$ are independent copies of a gaussian stochastic
stationary process $f_{t}$. This means that only degrees of freedom
from the set $\Lambda^{(m)}$ are subjected to damping (defined by
the factor $\alpha>0$) and to the external forces $F_{t,k}$.

If $\alpha=0,f_{t}=0$, then the system is the linear hamiltonian
system with the quadratic hamiltonian 
\begin{equation}
H(\psi)=\frac{1}{2}\sum_{i=1}^{N}p_{i}^{2}+\frac{1}{2}\sum_{i,j}V(j,i)q_{i}q_{j}=\frac{1}{2}((\begin{array}{cc}
V & 0\\
0 & E
\end{array})\psi,\psi)_{2}\label{hamiltonian}
\end{equation}
Note that the Gibbs distribution 
\begin{equation}
Z^{-1}\exp(-\beta H)=Z^{-1}\exp(-\frac{1}{2}(C_{G,\beta}^{-1}\psi,\psi)_{2}),\label{Gibbs}
\end{equation}
corresponding to the hamiltonian (\ref{hamiltonian}), is gaussian,
and 
\begin{equation}
C_{G,\beta}=C_{Gibbs}=\frac{1}{\beta}(\begin{array}{cc}
V^{-1} & 0\\
0 & E
\end{array})\label{C_G}
\end{equation}
is its covariance matrix.

One can rewrite system (\ref{system_2N}) in the vector notation

\begin{equation}
\frac{d\psi}{dt}=A\psi+F{}_{t},\label{main_matrix_form}
\end{equation}
where 
\begin{equation}
A=(\begin{array}{cc}
0 & E\\
-V & -\alpha D
\end{array})\label{A}
\end{equation}
$E$ is the unit $(N\times N)$-matrix, and $D$ is the diagonal $(N\times N)$-matrix
with all zeroes on the diagonal except $D_{k,k}=1,k=N-m+1,...,N$.

\subsection{Classes of Hamiltonians}

For any $N$ let $\mathbf{H}_{N}$ denote the set of all hamiltonians
(\ref{hamiltonian}) with positive definite $V$. Note that the dimension
of this set is $\dim\mathbf{\mathbf{H}_{N}}=\frac{N(N+1)}{2}$, that
coincides with the dimension of the set of symmetric $V$. In fact,
take some symmetric positive definite $V$, for example diagonal,
then any $V+V_{1}$, where $V_{1}$ is symmetric and has sufficiently
small elements, will be positive definite.

More generally, let $\Gamma=\Gamma_{N}$ be connected graph with $N$
vertices $i=1,...,N$, and not more than one edge per each (unordered)
pair of vertices $(i,j)$. It is assumed that all loops $(i,i)$ are
the edges of $\Gamma$. Denote $\mathbf{H}_{\Gamma}$ the set of (positive
definite) $V$ such that $V(i,j)=0$ if $(i,j)$ is not the edge of
$\Gamma$. The same argument shows that the dimension of $\mathbf{H}_{\Gamma}$
is equal to the number of edges of $\Gamma$. Note that $\mathbf{H}_{N}=\mathbf{H}_{\Gamma}$
for the complete graph $\Gamma$ with $N$ vertices.

In particular, we can consider the $d$-dimensional integer lattice
$Z^{d}$ and the graph $\Gamma=\Gamma(d,\Lambda)$, the set of vertices
of which is the cube 
\[
\Lambda=\Lambda(d,M)=\{(x_{1},...,x_{d})\in Z^{d}:|x_{i}|\leq M,i=1,...,d\}\subset Z^{d}
\]
and the edges $(i,j),\,|i-j|\leq1$.

In general, $V$ is called $\gamma$-local on $\Gamma$ if $V(i,j)=0$
for all pairs $i,j$ having distance $r(i,j)$ between them greater
than $\gamma$, where the distance $r(i,j)$ between two vertices
$i,j$ on a graph is the minimal length (number of edges) of paths
between them.

We shall say that some property holds for almost any hamiltonian from
the set $\mathbf{H}_{\Gamma}$ if the set $\mathbf{H}_{\Gamma}^{(+)}$,
where the property holds, is open and everywhere dense. One can prove
in fact that the dimension of the set $\mathbf{H}_{\Gamma}^{(-)}=\mathbf{H}_{\Gamma}\setminus\mathbf{H}_{\Gamma}^{(+)}$
where it does not hold, is less than the dimension of $\mathbf{H}_{\Gamma}$
itself.

\subsection{Invariant subspaces}

Consider the following subset of $L$ 
\[
L_{-}=\{\psi\in L:\ H(e^{tA}\psi)\to0,\ t\to\infty\}\subset L
\]
We will need the following result. Let $e_{i},i=1,...,N,$ be column
$N$-vectors with zero components except $i$-th component equal to
$1$.

\begin{lemma}\label{lemma_subspaces}

$L_{-}$ is a linear subspace of $L$ and $L_{-}=\{(\begin{array}{c}
q\\
p
\end{array})\in L:\ q\in l_{V},p\in l_{V}\}$, where $l_{V}$ is the subspace of $R^{N}$, spanned by the vectors
$V^{k}e_{i},\, i=N-m+1,...,N;\ k=0,1,\ldots$. Moreover, $L_{-}$
and its orthogonal complement denoted by $L_{0},$ are invariant with
respect to the operator $A$.

\end{lemma}

The proof is identical to the proof of theorem 2.1 in \cite{LM_1}.

\begin{lemma}\label{lemma_e_tA}

The spectrum of the restriction $A_{-}$ of $A$ on the subspace $L_{-}$
belongs to the left half-plane, and as $t\to\infty$ 
\[
||e^{tA_{-}}||_{2}\to0
\]
exponentially fast,

\end{lemma}

It follows because by definition of $L_{-}$ and boundedness of $H$
from below we have $e^{tA}\psi\to0$ for any $\psi\in L_{-}$.

\begin{lemma}\label{lemma_L_0}

For almost any $H\in\mathbf{H}_{\Gamma}$ we have $\dim L_{0}=0$.

\end{lemma}

Proof. For given $V$ the subspace $L_{-}=L_{-}(m)$ depends on $\Lambda^{(m)}$.
If $m_{1}<m_{2}$ then $L_{-}(m_{1})\subseteq L_{-}(m_{2})$. That
is why it is sufficient to prove the lemma in case of one-point subset
$\Lambda^{(1)}$. If $l_{V}$ is spanned by the vectors $V^{k}e_{N},k=0,1,...$,
then it is spanned by $N$ vectors $V^{k}e_{N},k=1,...,N$, and obviously
vice-versa. Let $\Sigma(V)$ be the $(N\times N)$-matrix the columns
of which are the vectors $V^{k}e_{N},k=1,...,N$. Thus, the inequality
$\det(\Sigma(V))\ne0$ for matrix $V\in\mathbf{H}_{\Gamma}$ is equivalent
to the statement that the vectors $V^{k}e_{N},k=1,...,N$ are linearly
independent, or $\dim l_{V}=N$. Then the set $\mathbf{H}_{\Gamma}^{(-)}$
of hamiltonians for which $\dim L_{0}>0$ is 
\[
\mathbf{H}_{\Gamma}^{(-)}=\{V:\ \dim(l_{V})<N\}=\{V:\det(\Sigma(V))=0\}
\]
Thus, $\mathbf{H}_{\Gamma}^{(-)}$ is the set of zeros of polynomial
function on a smooth manifold $\mathbf{H}_{\Gamma}$. Thus its dimension
is less than the dimension of $\mathbf{H}_{\Gamma}$.

\subsection{Covariances}

All our external forces $f_{t}$ will be gaussian stationary processes
with zero mean. Among them there is the white noise - the generalized
stationary gaussian process having covariance $C_{f}(s)=\sigma^{2}\delta(s)$,
it is sometimes called process with independent values (without memory).
All other stationary gaussian processes, which we consider here, are
processes with memory. We will assume that they have continuous trajectories
and integrable (short memory) covariance 
\[
C_{f}(s)=<f_{t}f_{t+s}>
\]
Then the solution of (\ref{main_matrix_form}) with arbitrary initial
vector $\psi(0)$ is unique and is equal to (for the white noise case
see for example \cite{Ventsel}, section 12.4) 
\begin{equation}
\psi(t)=e^{tA}(\int_{0}^{t}e^{-sA}F{}_{s}ds+\psi(0))\label{solution_psy_t}
\end{equation}

Our goal is to show that even weak memory, in the generic situation,
prevents the limiting invariant measure (which always exists and unique)
from being Gibbs. To formulate more readable results we assume more:
$C_{f}$ belongs to the Schwartz space $S=S(R)$. Then also the spectral
density 
\[
a(\lambda)=\frac{1}{2\pi}\int_{-\infty}^{+\infty}e^{-it\lambda}C_{f}(t)\ dt
\]
belongs to the space $S$.

We shall say that some property (for given $V$) holds for almost
all $C_{f}$ from the space $S$ if the set $S^{(+)}\subset S$ where
this property holds is open and everywhere dense in $S$.

\section{Main Results}

\subsection{Finite system}

Further on we denote, using (\ref{C_G}), 
\[
C_{G}=C_{G,2\alpha}
\]
Fix some connected graph $\Gamma$ with $N$ vertices.

\begin{theorem}\label{th_1}

Let $f_{t}$ be either white noise or has continuous trajectories
and integrable $C_{f}$. Then for any hamiltonian $H\in\mathbf{H}_{\Gamma}$
with $L_{0}=L_{0}(H)=\{0\}$ the following holds: 
\begin{enumerate}
\item there exists gaussian random $(2N)$-vector $\psi(\infty)$ such that
for any initial condition $\psi(0)$ the distribution of $\psi(t)$
converges, as $t\to\infty$, to that of $\psi(\infty)$; 
\item for the covariance of the process $\psi(t)$ we have $E\psi(t)\to0$
and 
\begin{equation}
C_{\psi(\infty)}(s)=\lim_{t\rightarrow\infty}<\psi(t)\psi^{T}(t+s)>=\lim_{t\rightarrow\infty}C_{\psi}(t,t+s)=W(s)C_{G}+C_{G}W(-s)^{T},\label{C_psi_t_plus_s}
\end{equation}
where 
\begin{equation}
W(s)=\int_{0}^{+\infty}e^{\tau A}C_{f}(\tau+s)d\tau\label{W}
\end{equation}

\end{enumerate}
\end{theorem}

\begin{corollary}\label{Cor_1}

For the white noise with variance $\sigma^{2}$ the vector $\psi(\infty)$
has Gibbs distribution (\ref{Gibbs}) with the temperature 
\[
\beta^{-1}=\frac{\sigma^{2}}{2\alpha}
\]

\end{corollary}

For $m=1$ this corollary was proved in \cite{LM_2}. Denote $C_{\psi}=C_{\psi(\infty)}(0)$.
Further on we denote the matrix elements of the matrix $C_{\mu}$
(and other $(2N)\times(2N)$-matrices as well) as for example $C_{\mu}(i,N+j)=C_{\mu}(q_{i},p_{j})$.

\begin{theorem}\label{th_2}

Let $N\geq2$, fix some graph $\Gamma$ and any $H\in\mathbf{H}_{\Gamma}$
with $L_{0}=L_{0}(H)=\{0\}$. Then the following assertions hold: 
\begin{enumerate}
\item for any $C_{f}\in S$ in the limiting distribution there are no correlations
between coordinates and velocities, that is $C_{\psi}(q_{i},p_{j})=0$
for any $i,j$; 
\item for almost any $C_{f}\in S$ there are nonzero correlations between
velocities, that is for some $i\neq j$ $C_{\psi}(p_{i},p_{j})\neq0$.
Thus, the limiting distribution cannot be Gibbs; 
\end{enumerate}
\end{theorem}

\subsection{Large $N$}

It is more interesting, however, that the convergence to Gibbs is
impossible even in the points of $\Lambda$ far away from the boundary,
in the thermodynamic limit $N\to\infty$.

The following result reduces (for large $N)$ calculation of the matrix
$C_{\psi}$ to that of the simpler matrix 
\[
C_{V}=\frac{\pi}{\alpha}(\begin{array}{cc}
a(\sqrt{V})V^{-1} & 0\\
0 & a(\sqrt{V})
\end{array})
\]
where $\sqrt{V}$ is the unique positive root of $V$.

\begin{remark}It is interesting to note that: 1) $C_{V}$ also defines
an invariant measure with respect to pure (that is with $\alpha=0,F_{t}=0$)
hamiltonian dynamics; 2) for the white noise case $C_{V}$ corresponds
to the Gibbs distribution.

\end{remark}

We assume that some graph $\Gamma$ is given with the set of vertices
$\Lambda,|\Lambda|=N,$ and the boundary set $\Lambda^{(m)}$. For
any $V\in\mathbf{H}_{\Gamma}$ such that $L_{0}(V)=\{0\}$, the following
representation of the limiting covariance matrix appears to be crucial
\[
C_{\psi}=C_{V}+Y_{V}
\]
where $Y_{V}$ is some remainder term. The following theorem gives
the estimates for $Y_{V}$. The norm $||V||_{\infty}$ of a matrix
V we define by the formula 
\[
||V||_{\infty}=\max_{i}\sum_{j}|V(i,j)|
\]

\begin{theorem} \label{th_3} Assume that $V$ is $\gamma$-local
and $||V||_{\infty}<B$ for some $B>0$. Fix also some number $\eta=\eta(N)\geqslant\gamma$.
The following assertion holds: 
\begin{enumerate}
\item If $C_{f}\in S$ and has bounded support, that is $C_{f}(t)=0$ if
$|t|>b$ for some $b>0$, then for any pair $i,j$ far away from the
boundary, that is on the distance $r(i,\Lambda^{(m)}),\ r(j,\Lambda^{(m)})>\eta(N)$,
there is the following estimate 
\[
|Y_{V}(q_{i},q_{j})|,\ |Y_{V}(p_{i},p_{j})|<|\Lambda^{(m)}|K_{0}\left(\frac{K}{\eta}\right)^{\eta\gamma^{-1}}
\]
for some constants $K_{0}=K(C_{f},B,b,\alpha,\gamma)$ and $K=K(C_{f},B,b,\alpha,\gamma)$,
not depending on $N$. 
\item For arbitrary $C_{f}\in S$ the estimate is 
\[
|Y_{V}(q_{i},q_{j})|,\ |Y_{V}(p_{i},p_{j})|<|\Lambda^{(m)}|C(k)\eta^{-k},
\]
for any $k>0$ and some constant $C(k)=C(C_{f},k,B,\alpha,\gamma)$,
not depending on $N$. 
\end{enumerate}
\end{theorem}

This theorem allows to do various conclusions concerning the thermodynamic
limit. We give an example.

For example, fix some $C_{f}(t)\in S$ and some connected countable
graph $\Gamma_{\infty}$ with the set of vertices $\Lambda_{\infty}$
and an increasing sequence of subsets $\Lambda_{1}\subset\Lambda_{2}\subset...\subset\Lambda_{n}\subset...$
such that $\Lambda=\cup\Lambda_{n}$. Let $\Gamma_{n}$ be the subgraph
of $\Gamma_{\infty}$ with the set of vertices $\Lambda_{n}$, that
is $\Gamma_{n}$ inherits all edges between vertices of $\Lambda_{n}$
from $\Gamma$. Denote $N_{n}=|\Lambda_{n}|$ and assume that the
boundaries $\Lambda_{n}^{(m)}$ are given with $m=m(n)$ such that
the following conditions holds: 
\begin{enumerate}
\item there exists $d>0$ such that for any $i\in\Lambda_{\infty}$ there
exists $n(i)$ such that for any $n>n(i)$ the following inequality
holds 
\[
r_{n}(i,\Lambda_{n}^{(m)})>\max\{m(n)^{\frac{1}{d}},\gamma\},
\]
where $r_{n}(i,\Lambda_{n}^{(m)})$ is the distance from vertex $i$
to the boundary $\Lambda_{n}^{(m)}$ on the graph $\Gamma_{n}$, 
\item for any $i\in\Lambda_{\infty}$ we have $r_{n}(i,\Lambda_{n}^{(m)})\rightarrow\infty$
as $n\rightarrow\infty$ (that is the boundary runs to infinity with
$n$). 
\end{enumerate}
Let $l^{\infty}(\Gamma_{\infty})$ be the complex Banach space of
bounded functions on the set of vertices of $\Gamma_{\infty}$: 
\[
l^{\infty}(\Gamma_{\infty})=\{(x_{i})_{i\in\Gamma_{\infty}}:\ \sup_{i\in\Gamma_{\infty}}|x_{i}|<\infty,\ x_{i}\in\mathbb{C}\}
\]
Fix some $\gamma$-local infinite matrix $V$ on this space and such
that $||V||_{\infty}\leqslant B$. It is clear that $V$ defines a
bounded linear operator on $l^{\infty}(\Gamma_{\infty})$. Denote
$\sigma(V)$ the spectrum of this operator. Let $V_{n}=(V(i,j))_{i,j\in\Lambda_{n}}$
be the restriction of $V$ on $\Lambda_{n}$, it is a matrix of the
order $N_{n}$. Assume that for all $n=1,2,\ldots$ the matrices $V_{n}$
are positive-definite. Note that the condition $L_{-}(V_{n})=L$ may
not hold for some $n$. However, one can choose a sequence of positive-definite
matrices $V'_{n}\in\mathbf{H}_{\Lambda_{n}}$ suh that $||V_{n}-V'_{n}||_{\infty}\to0$
as $n\to\infty$ with $L_{0}(V_{n}')=\{0\}$. Moreover, the convergence
of $V'_{n}$ to $V{}_{n}$ can be chosen arbitrary fast. Denote $C_{\psi}^{(n)}$
the limiting covariance matrices corresponding to $V'_{n}$.

\begin{corollary} \label{cor_2} The following assertions hold: 
\begin{enumerate}
\item for any $i,j\in\Lambda_{\infty}$ there exists the thermodynamic limit
\[
\lim_{n\to\infty}C_{\psi}^{(n)}(p_{i},p_{j})=C_{\psi}^{(\infty),p}(i,j),
\]
that is for distribution of velocities; 
\item if for any $i,j\in\Lambda_{\infty}$ there exists finite limits :
\begin{equation}
U(i,j)\doteqdot\lim_{n\rightarrow\infty}V_{n}^{-1}(i,j),\label{0605131}
\end{equation}
then for the coordinates we have 
\[
\lim_{n\to\infty}C_{\psi}^{(n)}(q_{i},q_{j})=C_{\psi}^{(\infty),q}(i,j)
\]

\item assume that the spectral density $a(\sqrt{\lambda})$ is analytic
on the open set containing the spectrum $\sigma(V)$. Then 
\[
C_{\psi}^{(\infty),p}(i,j)=a(\sqrt{V}),
\]
where $a(\sqrt{V})$ is defined in terms of the operator calsulus
on $l^{\infty}(\Gamma_{\infty})$ (\cite{Danford}, p. 568). 
\end{enumerate}
\end{corollary}

Let us add some comments to this corollary. Firstly, we want to emphasize
that in point 2 there are no any restrictions on $U(i,j)$. Secondly,
it is easy to see that the condition of point 3 is is fullfilled if
$C_{f}$ has bounded support (in this case the spectral is an entire
function). And finally, the thermodynamic limit typically is not Gibbs,
more exactly $C_{\xi}^{(\infty),p}(i,j)\ne0$ for any $i\neq j$ in
$\Lambda_{\infty}$ such that $a(\sqrt{V})(i,j)\neq0$.

\section{Proof of theorem 1}

The process $\psi(t)$ is not stationary. However, the following calculation
shows that it is asymptotically stationary.

Let $D^{(2)}$ be the diagonal $(2N\times2N)$-matrix with all zero
elements on the diagonal except $D_{k,k}^{(2)}=1,k=2N-m+1,...,2N$.
Obviously $D^{(2)}=(D^{(2)})^{T}=D^{(2)}(D^{(2)})^{T}$.

Then

\[
C_{\psi}(t,t+s)=E\int_{0}^{t}dt_{1}e^{(t-t_{1})A}F_{t_{1}}\int_{0}^{t+s}F_{t_{2}}^{T}e^{(t+s-t_{2})A^{T}}dt_{2}=
\]
\begin{equation}
=e^{tA}\int_{0}^{t}dt_{1}e^{-t_{1}A}D_{2}D_{2}^{T}\int_{0}^{t+s}dt_{2}e^{-t_{2}A^{T}}C_{f}(t_{1}-t_{2})e^{(t+s)A^{T}}\label{main_double_integral}
\end{equation}

For better understanding the following calculations, it is useful
to start with the white noise case, i. e. when 
\[
C_{f}(s)=\sigma^{2}\delta(s)
\]
It is a generalized function but the calculation follows the same
line. For $s=0$ (\ref{main_double_integral}) becomes 
\[
\sigma^{2}\int_{0}^{t}dt_{1}e^{(t-t_{1})A}D^{(2)}e^{(t-t_{2})A^{T}}
\]
We use a straightforward algebraic calculation with $(2\times2)$-block
matrices (\ref{C_G}) and (\ref{A}) to get 
\begin{equation}
AC_{G}+C_{G}A^{T}=-D^{(2)}\label{g_g_T}
\end{equation}
where $C_{G}$ is given by (\ref{C_G}) with $\beta=2\alpha$. Then
\begin{equation}
\frac{d}{dt}(e^{-tA}C_{G}e^{-tA^{T}})=e^{-tA}D^{(2)}e^{-tA^{T}}\label{derivative_eCe}
\end{equation}
and thus 
\[
C_{\psi}(t,t)=\sigma^{2}e^{tA}(\int_{0}^{t}dt_{1}e^{-t_{1}A}D^{(2)}e^{-t_{1}A^{T}})e^{tA^{T}}=\sigma^{2}e^{tA}(e^{-tA}C_{G}e^{-tA^{T}}-C_{G})e^{tA^{T}}
\]
and as $t\to\infty$ 
\[
C_{\psi}(t,t)\to_{t\to\infty}\sigma^{2}C_{G}
\]
This proves Corollary 1. Similarly one can show that $W(s)=0,s>0$,
$W(0)=\frac{\sigma^{2}}{2}E$ and $W(s)=\frac{1}{2}\sigma^{2}e^{-sA},s<0$.

In the general case define the new variables $t'_{i}=t-t_{i},i=1,2$.
Then the integral can be rewritten as 
\[
\int_{0}^{t}dt'_{1}e^{t'_{1}A}D^{(2)}\int_{-s}^{t}dt'_{2}e^{t'_{2}A^{T}}C_{f}(t'_{1}-t'_{2})e^{sA^{T}}
\]
Now we see that the limit $t\to\infty$ exists (first assertion of
theorem 1) and we can write it, using Lemma \ref{lemma_e_tA}, as
\[
\int_{0}^{\infty}dt'_{1}e^{t'_{1}A}D^{(2)}\int_{-s}^{\infty}dt'_{2}e^{t'_{2}A^{T}}C_{f}(t'_{1}-t'_{2})e^{sA^{T}}
\]
First consider the case $s=0$. We integrate over the quarter plane
$t'_{1}\geq0,t'_{2}\geq0$. Put $t'_{1}=t'_{2}+\tau$. Consider two
cones $\tau>0$ and $\tau<0$. Integration over the first (lower),
using gives 
\[
\int_{\tau>0}d\tau\int_{\tau}^{\infty}dt'_{1}e^{t'_{1}A}D^{(2)}e^{t'_{1}A^{T}}e^{-\tau A^{T}}C_{f}(\tau)=\int_{\tau>0}e^{\tau A}C_{G}e^{\tau A^{T}}e^{-\tau A^{T}}C_{f}(\tau)d\tau=
\]
\[
=\int_{\tau>0}e^{\tau A}C_{G}C_{f}(\tau)d\tau
\]
Symmetrically, integration over the upper angle gives 
\[
\int_{\tau>0}C_{G}e^{\tau A^{T}}C_{f}(\tau)d\tau
\]
The case $s>0$ is considered similarly. We have

\[
\int_{0}^{\infty}dt'_{1}e^{t'_{1}A}D^{(2)}\int_{-s}^{\infty}dt'_{2}e^{t'_{2}A^{T}}C_{f}(t'_{1}-t'_{2})e^{sA^{T}}
\]
We integrate over the quarter plane $t'_{1}\geq0,t'_{2}\geq-s$. Put
$t'_{1}=t'_{2}+\tau$. The domain of integration $(\tau,t'_{1})$
cosists of two non-intersecting subdomains: the first one is a \textquotedbl{}shifted\textquotedbl{}
quarter-plane $\Omega_{1}=\{(\tau,t'_{1}):\ \tau<s,\ t'_{1}>0\}$,
the second is the cone $\Omega_{2}=\{(\tau,t'_{1}):\ \tau>s,\ t'_{1}>\tau-s\}$.
For the integral over $\Omega_{1}$ we have 
\[
\int_{\tau<s}d\tau\int_{0}^{\infty}dt'_{1}e^{t'_{1}A}D^{(2)}e^{t'_{1}A^{T}}e^{-\tau A^{T}}C_{f}(\tau)\ e^{sA^{T}}=\int_{\tau<s}C_{G}e^{-\tau A^{T}}C_{f}(\tau)d\tau\ e^{sA^{T}}=
\]
\[
=C_{G}\int_{\tau<s}e^{-\tau A^{T}}C_{f}(\tau)d\tau\ e^{sA^{T}}.
\]
Changing variables $\tau'=s-\tau$ we have 
\[
\int_{\tau<s}e^{-\tau A^{T}}C_{f}(\tau)d\tau\ e^{sA^{T}}=\int_{0}^{+\infty}e^{\tau'A^{T}}C_{f}(\tau'-s)d\tau'=W^{T}(-s)
\]
The integral over the cone gives 
\[
\int_{\tau>s}d\tau\int_{\tau-s}^{+\infty}dt'_{1}e^{t'_{1}A}D^{(2)}e^{t'_{1}A^{T}}e^{-\tau A^{T}}C_{f}(\tau)\ e^{sA^{T}}=\int_{\tau>s}e^{(\tau-s)A}C_{G}e^{(\tau-s)A^{T}}e^{-\tau A^{T}}C_{f}(\tau)d\tau\ e^{sA^{T}}
\]
\[
=\int_{\tau>s}e^{(\tau-s)A}C_{G}C_{f}(\tau)d\tau=\int_{\tau>0}e^{\tau A}C_{f}(\tau+s)d\tau C_{G}=W(s)C_{G}
\]

\section{Proof of Theorem 2}

We will need another expression for $C_{\psi}$ - in terms of the
spectral density of the process $f_{t}$ and the resolvent of $A$
\[
R_{A}(z)=(A-z)^{-1}
\]

\begin{lemma}

Fix any $C_{f}\in S$. Then for almost any $H\in\mathbf{H}_{\Gamma}$
the following assertions hold:

\begin{equation}
C_{\psi}=-\int_{-\infty}^{+\infty}a(\lambda)(R_{A}(i\lambda)C_{G}+C_{G}R_{A}^{T}(i\lambda))d\lambda;\label{C_psi_a_lambda}
\end{equation}

\end{lemma}

To prove this we just express $W$ in terms of the spectral density
$a(\lambda)$ and the resolvent of $A$ 
\[
W=\int_{0}^{+\infty}C_{f}(s)e^{sA}ds=\int_{-\infty}^{+\infty}d\lambda\int_{0}^{+\infty}ds\ a(\lambda)e^{is\lambda}e^{sA}=
\]

\[
=-\int_{-\infty}^{+\infty}a(\lambda)(A+i\lambda)^{-1}d\lambda=-\int_{-\infty}^{+\infty}a(\lambda)R_{A}(i\lambda)d\lambda,
\]
where the symmetry of the spectral density $a(\lambda)=a(-\lambda)$
is used.

Explicit expressions for the matrix elements $C_{\psi}(q_{i},p_{j})$
of $C_{\psi(\infty)}(0)$ seem to be ugly. Instead we will write the
matrix $C_{\psi}$ in the two-block form. For example, 
\begin{equation}
R_{A}(z)C_{G}=\frac{1}{2\alpha}(\begin{array}{cc}
Q_{11} & Q_{12}\\
Q_{21} & Q_{22}
\end{array}),\label{R_A_C_G}
\end{equation}
where the $(N\times N)$-blocks $Q_{11},Q_{22},Q_{12}+Q_{21}^{T}$
give, after integration, the matrix elements $C_{\psi}(q_{i},q_{j}),C_{\psi}(p_{i},p_{j}),C_{\psi}(q_{i},p_{j})$
correspondingly.

To get explicit expression for $Q_{ij}$ we need some notation. Define
the following rational matrices: $N\times N$-matrices 
\[
\rho(z)=(V+z^{2})^{-1},\theta(z)=\rho(z)T(z)\rho(z),\,\, T(z)=\alpha\hat{e}\tau^{-1}(z)\hat{e}^{T}
\]
where 
\[
\tau(z)=E+\alpha z\kappa(z)
\]
is $m\times m$-matrix, $E=E^{(m)}$ is the unit $m\times m$-matrix,
$\hat{e}$ is the $(N\times m)$-matrix with the only non-zero entries
$\hat{e}_{N-m+i,i}=1,i=1,...,m$, and

\[
\kappa(z)=(\rho(z)_{i,j})_{i,j=N-m+1,\ldots,N}
\]
is the restriction of $\rho$ on $\Lambda^{(m)}$. It is clear that
$\kappa(z)=\hat{e}^{T}\rho(z)\hat{e}$.

\begin{lemma}\label{lemma_Q_ij}

The block matrices $Q_{ij}$ are given by 
\[
Q_{11}=-z\rho(z)V^{-1}-\theta(z),Q_{12}=-\rho(z)+z\theta(z),
\]
\begin{equation}
Q_{21}=zQ_{11}+V^{-1},Q_{22}=zQ_{12}\label{L0107GX}
\end{equation}

\end{lemma}

Multiplying left and right sides of (\ref{R_A_C_G}) on $A-zE$, we
get 4 equations for $(N\times N)$-matrices 
\begin{align}
V^{-1}= & \ Q_{21}-zQ_{11},\label{L01071}\\
0= & \ VQ_{11}+(\alpha D+zE)Q_{21},\label{L01072}\\
0= & \ Q_{22}-zQ_{12},\label{L010712}\\
-E= & \ VQ_{12}+(\alpha D+zE)Q_{22}.\label{L010713}
\end{align}
It is clear that (\ref{L01071}) and (\ref{L010712}) are equivalent
to the first and second equalities (\ref{L0107GX}) correspondingly.
Note also the following simple equality 
\begin{equation}
z\alpha D\rho T=\alpha^{2}z\hat{e}\ (\hat{e}^{T}\rho\hat{e})\ \tau^{-1}\hat{e}^{T}=\alpha^{2}z\hat{e}\kappa(z)\tau^{-1}\hat{e}^{T}=\alpha\hat{e}(\tau-E)\tau^{-1}\hat{e}^{T}=\alpha D-T.\label{L0107DRD}
\end{equation}
where $D$ -is the diagonal $N\times N)$-matrix, introduced above
as the projection onto the subspace generated by the vectors $e_{N-m+1},\ldots,e_{N}$.
We get (\ref{L010713}), expressing $Q_{22}$ through $Q_{12}$, using
the second equality (\ref{L0107GX}), 
\[
VQ_{12}+(\alpha D+zE)Q_{22}=(V+z^{2})Q_{12}+z\alpha DQ_{12}=
\]
\[
=-E+zT\rho-z\alpha D\rho+z(z\alpha D\rho T)\rho=-E+zT\rho-z\alpha D\rho+z(\alpha D-T)\rho=-E.
\]
Thus, we have proved (\ref{L010713}). Note that the following equality
holds 
\[
z^{2}\rho V^{-1}=(z^{2}+V)\rho V^{-1}-V\rho V^{-1}=V^{-1}-\rho
\]

Similarly, check (\ref{L01072}), expressing $Q_{21}$ through $Q_{11}$,
using (\ref{L0107GX}), 
\[
VQ_{11}+(\alpha D+zE)Q_{21}=(V+z^{2})Q_{11}+z\alpha DQ_{11}+(\alpha D+zE)V^{-1}=
\]

\[
=-zV^{-1}-(V+z^{2})\theta+z\alpha D(-z\rho V^{-1}-\theta)+(\alpha D+zE)V^{-1}=
\]
\[
=-(V+z^{2})\theta-z^{2}\alpha D\rho V^{-1}-z\alpha D\theta+\alpha DV^{-1}=-(V+z^{2})\theta-\alpha D(V^{-1}-\rho)-z\alpha D\theta+\alpha DV^{-1}=
\]
\[
=-(V+z^{2})\theta+\alpha D\rho-z\alpha D\theta=-T\rho+\alpha D\rho-(z\alpha D\rho T)\rho=-T\rho+\alpha D\rho-(\alpha D-T)\rho=0.
\]
Lemma is proved.

Now we will prove theorem 2. To prove the first part it is sufficient
to take the sum of $R_{A}(z)C_{G}$ and its transposition. that is
to verify that $Q_{12}+Q_{21}=0$. But it is a simple calculation
using Lemma \ref{lemma_Q_ij}.

As for the second part of theorem 2, we should show that the $(N\times N)$-matrix
equation for lower diagonal block 
\begin{equation}
-\frac{1}{\alpha}\int_{-\infty}^{+\infty}a(\lambda)Q_{22}(i\lambda)d\lambda=E\label{Q_22_E}
\end{equation}
is rarely fulfilled. One can see that the matrix elements of $Q_{22}$
are bounded because matrix elements of the resolvent $R_{A}(i\lambda)$
are bounded, and moreover have no poles by lemma \ref{lemma_e_tA}.
This is not clear from the explicit expression 
\[
Q_{22}(z)=-z\rho(z)+z^{2}\rho(z)T(z)\rho(z)=z\rho(z)(-E+zT(z)\rho(z))=
\]
\[
=z\rho(z)(-1+\alpha z\hat{e}(E+\alpha z\kappa(z))^{-1}\hat{e}^{T}\rho(z))
\]
Equation (\ref{Q_22_E}) is equivalent to $N^{2}$ equations with
respect to the function $a(\lambda)$, given $V$. Each of these equations
is of the type 
\begin{equation}
\int_{-\infty}^{\infty}a(\lambda)\varphi(V(i,j),\lambda)d\lambda=0\label{Gibbs-nonGibbs}
\end{equation}
for some bounded function $\varphi$. The set of solutions is a closed
subset of the Schwartz space and that in a any small neighborhood
of any solution (of even one of the equations) there is open set of
points which do not satisfy this equation. Thus, the complement is
an everywhere dense subset. It is an open subset because if some $a(\lambda)$
does not satisfy the equation then its small neighborhood also not.

\section{Large $N$ - proof of theorem 3}

We will find now the main term of $C_{\psi}$ for large $N$. Decompose
matrix $A$ as follows 
\[
A=A_{V}+A_{D}
\]
\[
A_{V}=(\begin{array}{cc}
0 & E\\
-V & 0
\end{array}),A_{D}=(\begin{array}{cc}
0 & 0\\
0 & -\alpha D
\end{array})
\]
and use the formula 
\begin{equation}
e^{t(A_{V}+A_{D})}=e^{tA_{V}}+Y_{t}\label{exp_tA}
\end{equation}
where 
\begin{equation}
Y_{t}=\int_{0}^{t}e^{(t-s)A_{V}}A_{D}e^{sA}\ ds\label{R_t}
\end{equation}

Then by theorem \ref{th_1} for any $b>0$ we can write 
\begin{equation}
C_{\psi}=W(0)C_{G}+C_{G}W(0)^{T}=C_{V}+Y_{V,b}+Y_{V,\infty},\label{C_V_plus_R}
\end{equation}
where 
\[
C_{V}=\int_{0}^{+\infty}C_{f}(s)e^{sA_{V}}ds\ C_{G}+C_{G}\int_{0}^{+\infty}C_{f}(s)e^{sA_{V}^{T}}ds
\]
\[
Y_{V,b}=\int_{0}^{b}C_{f}(s)Y_{s}ds\ C_{G}+C_{G}\int_{0}^{b}C_{f}(s)Y_{s}^{T}ds
\]
\[
Y_{V,\infty}=\int_{b}^{\infty}C_{f}(s)Y_{s}ds\ C_{G}+C_{G}\int_{b}^{\infty}C_{f}(s)Y_{s}^{T}ds
\]

First we will find $C_{V}$.

\begin{lemma}\label{lemma_C_V_a} We have 
\[
C_{V}=\frac{\pi}{\alpha}(\begin{array}{cc}
a(\sqrt{V})V^{-1} & 0\\
0 & a(\sqrt{V})
\end{array}),
\]
\end{lemma}

Proof. Using the formula (see, for example \cite{DalKrein}, section
II.3) 
\begin{equation}
\exp(tA_{V})=(\begin{array}{cc}
\cos(\sqrt{V}t) & (\sqrt{V})^{-1}\sin(\sqrt{V}t)\\
-\sqrt{V}\sin(\sqrt{V}t) & \cos(\sqrt{V}t)
\end{array}).\label{062513}
\end{equation}
one can get 
\[
C_{V}=\frac{1}{\alpha}\int_{0}^{+\infty}C_{f}(s)(\begin{array}{cc}
V^{-1}\cos(\sqrt{V}s) & 0\\
0 & \cos(\sqrt{V}s)
\end{array})\ ds.
\]
Let $dE_{\lambda}$ be the spectral presentation for $V$, then 
\[
V=\int\lambda dE_{\lambda},\,\,\cos(\sqrt{V}s)=\int_{-\infty}^{+\infty}\cos(\sqrt{\lambda}s)dE_{\lambda}.
\]
where the integral is taken only over positive half-axis because of
the spectrum of $V$. Thus 
\[
\int_{0}^{+\infty}C_{f}(s)\cos(\sqrt{V}s)ds=\int_{-\infty}^{+\infty}(\int_{0}^{+\infty}C_{f}(s)\cos(\sqrt{\lambda}s)ds)dE_{\lambda})=\pi\int_{-\infty}^{+\infty}a(\sqrt{\lambda})dE_{\lambda}=\pi a(\sqrt{V}).
\]

Lemma is proved.

Now we will prove theorem \ref{th_3} for the case when $a(\lambda)$
has bounded support $[-b,b]$. Let us estimate matrix elements of
\[
Y_{V,b}=\int_{0}^{b}C_{f}(s)Y_{s}\ ds\ C_{G}+C_{G}\int_{0}^{b}C_{f}(s)Y_{s}^{T}\ ds,
\]
where 
\[
Y_{t}=\int_{0}^{t}e^{(t-s)A_{V}}A_{D}e^{sA}\ ds.
\]
Denote $U_{i,j}(s,t)=(e^{(t-s)A_{V}}A_{D}e^{sA}\ C_{G})(p_{i},p_{j})$.
Then 
\[
U_{i,j}(s,t)=\sum_{k_{1},k_{2},k_{3}}\sum_{x_{k_{1}},x_{k_{2}},x_{k_{3}}}e^{(t-s)A_{V}}(p_{i},x_{k_{1}})A_{D}(x_{k_{1}},x_{k_{2}})e^{sA}(x_{k_{2}},x_{k_{3}})C_{G}(x_{k_{3}},p_{j}).
\]
where $x_{k}$ can be either $q_{k}$ or $p_{k}$. It is clear that
the terms of this sum can be non-zero only if $x_{k_{3}}=p_{j}$ and
$x_{k_{1}}=x_{k_{2}}=p_{k}$, where $k\in\Lambda^{(m)}$. Thus 
\begin{equation}
U_{i,j}(s,t)=-\frac{1}{2}\sum_{k\in\Lambda^{(m)}}e^{(t-s)A_{V}}(p_{i},p_{k})e^{sA}(p_{k},p_{j}).\label{0625132}
\end{equation}
\begin{lemma} For any $k\in\Lambda^{(m)}$ we have 
\[
|e^{(t-s)A_{V}}(p_{i},p_{k})|\leqslant\frac{(\sqrt{B}(t-s))^{r(i)}}{r(i)!}e^{\sqrt{B}(t-s)}
\]
where $r(i)=2[\gamma^{-1}r(i,\Lambda^{(m)})]$ is an integer. \end{lemma}

We have 
\begin{equation}
e^{(t-s)A_{V}}(p_{i},p_{k})=\cos(\sqrt{V}(t-s))(i,k)=\sum_{n=0}^{\infty}(-1)^{n}\frac{(t-s)^{2n}}{(2n)!}(V^{n})(i,k).\label{L0205}
\end{equation}
By locality of the hamiltonian $V$ we have $V^{n}(i,k)=0$, if $n<\gamma^{-1}r(i,\Lambda^{(m)})$.
Then (\ref{L0205}) can be estimated as 
\[
|e^{(t-s)A_{V}}(p_{i},p_{k})|=|\sum_{n=[\gamma^{-1}r(i,\Lambda^{(m)})]}^{\infty}(-1)^{n}\frac{(t-s)^{2n}}{(2n)!}(V^{n})(i,k)|\leqslant\sum_{n=[\gamma^{-1}r(i,\Lambda^{(m)})]}^{\infty}\frac{(t-s)^{2n}}{(2n)!}B^{n}\leqslant
\]
\[
\leqslant\frac{(\sqrt{B}(t-s))^{r(i)}}{(r(i))!}e^{\sqrt{B}(t-s)}.
\]
\begin{lemma} For any $k\in\Lambda^{(m)}$

\[
|e^{sA}(p_{k},p_{j})|\leqslant\frac{(cs)^{r(j)}}{r(j)!}e^{cs},
\]
where $r(j)=2[\gamma^{-1}r(j,\Lambda^{(m)})]$ and $c=B+\alpha$.

\end{lemma}

Consider the following expansion 
\[
e^{sA}=\sum_{n=0}^{\infty}\frac{s^{k}}{k!}A^{n}.
\]
Then 
\[
|A^{n}(p_{k},p_{j})|\leqslant||A^{n}||\leqslant||A||^{n}\leqslant c^{n},
\]
where $c=B+\alpha$. Moreover, let us prove that $A^{n}(p_{k},p_{j})=0$
for any $n$ such that 
\[
n<2\gamma^{-1}r(j,\Lambda^{(m)}).
\]
It is easy to see that $A^{n}(p_{k},p_{j})=(A_{V}+A_{D})^{n}(p_{k},p_{j})$
is the sum of the terms 
\[
(-\alpha)^{u_{0}}A_{V}^{u_{1}}(p_{k_{1}},p_{k_{2}})A_{V}^{u_{2}}(p_{k_{1}},p_{k_{2}})\ldots A_{V}^{u_{q}}(p_{k_{q}},p_{j}),\quad k_{1}=k
\]
where $k_{1},\ldots,k_{q}\in\Lambda^{(m)}$, $u_{0}+u_{1}+\ldots+u_{q}=n$
and $u_{l}\geqslant0$ for all $l=0,1,\ldots,q$. For the latter factor
we get, using (\ref{L0205}): 
\[
A_{V}^{u_{q}}(p_{k_{q}},p_{j})=\begin{cases}
(-1)^{u}V^{u}(k_{q},j), & u_{q}=2u,\\
0, & \mathrm{otherwise}
\end{cases}.
\]
By locality of $V$ we get that $A_{V}^{u_{q}}(p_{k_{q}},p_{j})=0$,
if $u<\gamma^{-1}r(j,\Lambda^{(m)})$. As $n\geqslant2u$, then $A^{n}(p_{k},p_{j})=0$
for all $n<2\gamma^{-1}r(j,\Lambda^{(m)})$. Then 
\[
|e^{sA}(p_{k},p_{j})|\leqslant\sum_{n=2[\gamma^{-1}r(j,\Lambda^{(m)})]}\frac{s^{n}}{n!}c^{n}\leqslant\frac{(cs)^{r(j)}}{r(j)!}e^{cs}
\]
and 
\[
|U_{i,j}(s,t)|\leqslant\frac{1}{2}|\Lambda^{(m)}|\frac{(B'(t-s))^{r(i)}(cs)^{r(j)}}{r(i)!r(j)!}e^{B'(t-s)+cs}\leqslant\frac{1}{2}|\Lambda^{(m)}|\frac{c_{1}^{r(i)+r(j)}}{r(i)!r(j)!}(t-s)^{r(i)}s^{r(j)}e^{c_{1}t},
\]
where $c_{1}=\sqrt{B}+c$. For the integral 
\[
|(Y_{t}C_{G})(p_{i},p_{j})|\leqslant\frac{1}{2}|\Lambda^{(m)}|\frac{c_{1}^{r(i)+r(j)}}{r(i)!r(j)!}e^{c_{1}t}\int_{0}^{t}(t-s)^{r(i)}s^{r(j)}ds=\frac{1}{2}|\Lambda^{(m)}|\frac{c_{1}^{r(i)+r(j)}}{(r(i)+r(j)+1)!}e^{c_{1}t}t^{r(i)+r(j)+1}=
\]
\[
=\frac{1}{2c_{1}}|\Lambda^{(m)}|\frac{(c_{1}t)^{r(i)+r(j)+1}}{(r(i)+r(j)+1)!}e^{c_{1}t}.
\]
and finally 
\[
|Y_{V,b}(p_{i},p_{j})|\leqslant|\Lambda^{(m)}|K_{0}\frac{(c_{1}b)^{r(i)+r(j)+1}}{(r(i)+r(j)+1)!}\leqslant|\Lambda^{(m)}|K_{0}\left(\frac{K}{\eta}\right)^{\gamma^{-1}\eta}
\]
where we have 
\begin{equation}
K=c_{1}b\gamma e\quad K_{0}=\frac{1}{c_{1}}C_{f}(0)e^{2c_{1}b}.\label{Constants_K_K_0}
\end{equation}
For $Y_{V,b}(q_{i},q_{j})$ the proof and the estimates are quite
similar and we omit the proof. The constant $K$ is the same as in
(\ref{Constants_K_K_0}) and the new constant $K_{0}$ is 
\[
\tilde{K}_{0}=\frac{1}{\sqrt{B}c_{1}}C_{f}(0)e^{2c_{1}\omega}.
\]

For arbitrary $C_{f}\in S$ the proof is as follows. Put $b=\sqrt{\eta}$
and estimate the integral over $(0,b)$ as above. Then we get: 
\begin{align*}
K_{0}= & \ \frac{1}{c_{1}}C_{f}(0)e^{2c_{1}\sqrt{\eta}}\leqslant\frac{1}{c_{1}}C_{f}(0)(e^{2c_{1}\gamma})^{\gamma^{-1}\eta},\\
|Y_{V,\omega}(p_{i},p_{j})|\leqslant & \ |\Lambda^{(m)}|\frac{1}{c_{1}}C_{f}(0)\left(\frac{c_{1}\gamma e^{2c_{1}\gamma+1}}{\sqrt{\eta}}\right)^{\gamma^{-1}\eta},\\
|Y_{V,\omega}(q_{i},q_{j})|\leqslant & \ |\Lambda^{(m)}|\frac{1}{\sqrt{B}c_{1}}C_{f}(0)\left(\frac{c_{1}\gamma e^{2c_{1}\gamma+1}}{\sqrt{\eta}}\right)^{\gamma^{-1}\eta}
\end{align*}
Then it is easy to see that for all $k=0,1,2,\ldots$ there exists
constant $C=C(c_{1},B,\gamma,k)$ such that any $\eta>0$ we have:
\[
|Y_{V,\omega}(p_{i},p_{j})|\leqslant|\Lambda^{(m)}|C\eta^{-k},\quad|Y_{V,\omega}(q_{i},q_{j})|\leqslant|\Lambda^{(m)}|C\eta^{-k}.
\]

To estimate the integral over $(b,\infty)$, we need the following
lemma.

\begin{lemma} For any $i,j\in\Lambda$ the following inequalities
hold 
\begin{align*}
|e^{tA}(p_{i},p_{j})| & \ \leqslant1,\quad|e^{tA}(q_{i},p_{j})|\leqslant t\\
|e^{tA_{V}}(p_{i},p_{j})| & \ \leqslant1,\quad\alpha|\left(e^{tA_{V}}C_{G}\right)(p_{i},q_{j})|\leqslant\frac{t}{2}
\end{align*}
\end{lemma}

Denote $e^{tA}g_{j}=(q(t),p(t))^{T}$, where $p(t)=(p_{1}(t),\ldots,p_{N}(t))^{T}$,
$q(t)=(q_{1}(t),\ldots,q_{N}(t))^{T}$, and the <<initial>> vector
$g_{j}=(0,e_{j})^{T}\in l^{(p)}$. From the definition of the matrix
exponent we have: 
\begin{align*}
e^{tA}(p_{i},p_{j})= & \ p_{i}(t),\\
e^{tA}(q_{i},p_{j})= & \ \int_{0}^{t}p_{i}(s)ds.
\end{align*}
Therefore the bound for $e^{tA}(q_{i},p_{j})$ in the lemma follows
from the inequality $|e^{tA}(p_{i},p_{j})|\leqslant1$. As the energy
along $e^{tA}g_{j}$ cannot increase we get: 
\begin{equation}
|e^{tA}(p_{i},p_{j})|^{2}=p_{i}^{2}(t)\leqslant2H((q(t),p(t))^{T})\leqslant2H((q(0),p(0))^{T})=1.\label{0529132}
\end{equation}
Thus, the inequalities for the matrix elements of $e^{tA}$ have been
proven. The estimate for $|e^{tA_{V}}(p_{i},p_{j})|$ is obtained
similarly (\ref{0529132}). Let us check the last inequality. From
the formula (\ref{062513}) we have: 
\[
e^{tA_{V}}C_{G}=\frac{1}{2\alpha}\left(\begin{array}{cc}
V^{-1}\cos(tR) & R^{-1}\sin(tR)\\
-R^{-1}\sin(tR) & \cos(tR)
\end{array}\right),
\]
where we have put $R=\sqrt{V}$. Thus we get: 
\begin{equation}
\alpha\left(e^{tA_{V}}C_{G}\right)(p_{i},q_{j})=-\frac{1}{2}\left(R^{-1}\sin(tR)\right)(i,j)\label{05171330}
\end{equation}
and 
\begin{equation}
\alpha|\left(e^{tA_{V}}C_{G}\right)(p_{i},q_{j})|\leqslant\frac{1}{2}||R^{-1}\sin(tR)||_{2}.\label{0529133}
\end{equation}
As $R$ is selfadjoint then 
\[
||R^{-1}\sin(tR)||_{2}=\max\{|\left|\frac{\sin t\lambda}{\lambda}\right|:\ \lambda\in\sigma(R)\}
\]
But for any $\lambda\in\mathbb{R}$ and $t\geqslant0$ we have 
\[
\left|\frac{\sin t\lambda}{\lambda}\right|\leqslant t,
\]
It follows $||R^{-1}\sin(tR)||_{2}\leqslant t$. Applying this estimate
for the norm to (\ref{0529133}), we get the final estimatefor $\alpha|\left(e^{tA_{V}}C_{G}\right)(p_{i},q_{j})|$.
Lemma is proved.

From this lemma and formula (\ref{0625132}) we get also the estimate
\[
|U_{ij}(s,t)|\leqslant\frac{1}{2}|\Lambda^{(m)}|
\]
for any $0\leqslant s\leqslant t$ and any $i,j$. Then 
\[
|Y_{V,\infty}(p_{i},p_{j})|\leqslant|\Lambda^{(m)}|\int_{\sqrt{\eta}}^{+\infty}s|C_{f}(s)|ds.
\]
By definition of the space $S$, it is clear from the last inequality
that for any $k>0$, the following inequality holds. 
\[
Y_{V,\infty}(p_{i},p_{j})\leq|\Lambda^{(m)}|C(k)\eta^{-k},\,\,\, C(k)=C(k,\alpha,B,\gamma).
\]
The estimate for coordinates can be proved similarly. Theorem is proved.

Let prove the corollary \ref{cor_2}. Further, $V_{n}'$ we write
$V_{n}$. From the previous theorem it follows that 
\begin{align}
\lim_{n\rightarrow\infty}|C_{\xi}^{(n)}(q_{i},q_{j})-C_{V_{n}}(q_{i},q_{j})|= & 0,\label{06051310}\\
\lim_{n\rightarrow\infty}|C_{\xi}^{(n)}(p_{i},p_{j})-C_{V_{n}}(p_{i},p_{j})|= & 0,\label{0604133}
\end{align}
Thus it is sufficient to show that the elements of the matrix $C_{V_{n}}$
have finite limits as $n\to\infty$.

\begin{lemma} \label{0604134} Let $P$ be an arbitrary polynom of
degree $d$, then for any $i,j\in\Gamma_{\infty}$

\[
\lim_{n\rightarrow\infty}P(V_{n})(i,j)=P(V)(i,j)
\]
\end{lemma} In fact, for example 
\[
|V^{2}(i,j)-V_{n}^{2}(i,j)|=|\sum_{k\notin\Lambda_{n}}V(i,k)V(k,j)|\leqslant||V||_{\infty}\sum_{k\notin\Lambda_{n}}|V(i,k)|\rightarrow0.
\]
as $n\rightarrow\infty$. And similarly for any degree of $V$.

\begin{lemma} \label{0604132} For any $i,j\in\Lambda_{\infty}$
there exists the finite limit 
\[
\lim_{n\rightarrow\infty}a(\sqrt{V_{n}})(i,j)=C_{\xi}^{(\infty),p}(i,j).
\]
\end{lemma} The function $a(\sqrt{x})$ is continuous on the segment
$[0,B]$, thus there exists a sequence of real polynomials $P_{k}(x),\ k=1,2,\ldots$
uniformly converging to $a(\sqrt{x})$ on $[0,B]$ as $k\rightarrow\infty$.
Note that the spectrum of $V_{n}$ belongs to $[0,B]$ for any $n=1,2,\ldots$.
Then the following inequalities hold 
\begin{equation}
|P_{k}(V_{n})(i,j)-a(\sqrt{V_{n}})(i,j)|\leqslant||P_{k}(V_{n})-a(\sqrt{V_{n}})||_{2}\leqslant\sup_{x\in[0,B]}|P_{k}(x)-a(\sqrt{x})|\label{0604131}
\end{equation}
The latter follows from the spectral mapping theorem (\cite{Danford},
p. 569). From (\ref{0604131}) it follows that $P_{k}(\sqrt{V_{n}})(i,j)\rightarrow a(\sqrt{V_{n}})(i,j)$
as $k\rightarrow\infty$, uniformly in $n=1,2,\ldots$. Then by lemma
\ref{0604134} we have the assertion of the lemma.

Let us now prove corollary \ref{cor_2}. The first poingt follows
immediately from the equality (\ref{0604133}) and lemma \ref{0604132}.
To prove the second assertion we use equality (\ref{06051310}). Rewrite
the elements $C_{V_{n}}(q_{i},q_{j})$ as 
\[
C_{V_{n}}(q_{i},q_{j})=\left(a(\sqrt{V_{n}})V_{n}^{-1}\right)(i,j)=\left(\left(a(\sqrt{V_{n}})-a(0)\right)V_{n}^{-1}\right)(i,j)+a(0)V_{n}^{-1}(i,j)=f(V_{n})(i,j)+a(0)V_{n}^{-1}(i,j),
\]
where we introduced the function $f(x)=(a(\sqrt{x})-a(0))x^{-1}$.
As the spectral density $a(x)$ is even, then $f(x)$ is continuous
on $\mathbb{R}_{\geqslant0}$. The arguments, similar to those in
the proof of lemma \ref{0604132}, show that for any $i,j\in\Lambda_{\infty}$
there exists the limit 
\[
\lim_{n\rightarrow\infty}f(V_{n})(i,j)
\]
As $V_{n}^{-1}(i,j)\rightarrow U(i,j)$ when $n\rightarrow\infty$,
the first two assertions of corollary 2 are proved.

The last assertion is similar to the proof of lemma \ref{0604132}
and lemma 13 in \cite{Danford}, p. 571, if applied to the sequence
$P_{k}(V)$.

\section{Comments}
\begin{enumerate}
\item For concrete $V$, even simply looking, it may be rather difficult
to find $\dim L_{0}$, and moreover, mostly it is not $0$. Example
is the one-dimensional harmonic chain 
\[
\sum_{i=-N}^{N}\omega_{0}q_{i}^{2}+\omega_{1}\sum_{i=-N}^{N-1}(q_{i}-q_{i+1})^{2}),\,\omega_{0},\omega_{1}>0
\]
where the calculation of $\dim L_{0}$ leads to number theory problems.
However, this dimension mostly is much less than the dimension of
$L$ itself (more exactly, is $o(N)$), see \cite{LM_1}. However,
one can always use instability of the integer $\dim L_{0}$: even
a smallest generic perturbation of $V$ leads to the desired zero
dimension effect. 
\item All questions concerning the alternative Gibbs-nonGibbs lead to equations
of the type (\ref{Gibbs-nonGibbs}). In theorem 2 we considered (\ref{Gibbs-nonGibbs})
as equation for $a(\lambda)$ with given $V$. However, one can ask
also the question dual to Theorem 2. Namely, fix arbitary $a(\lambda)\in S$,
is it true that for almost any $H\in\mathbf{H}_{G}$ there is a pair
$i\neq j$ such that $C_{\psi}(p_{i},p_{j})\neq0$. It is more or
less clear that the answer will be yes. We do not prove it carefully
here. For example, consider the famous Ornstein-Ulehnbeck process
with the spectral density 
\begin{equation}
a(\lambda)=\frac{c}{\mu^{2}+\lambda^{2}},\label{O_U_spectral}
\end{equation}
so that the limiting covariance had inter-velocity correlations for
a class of $V_{n}$ with $L_{0}=\emptyset$. It is easy to get such
examples. Assume that in (\ref{O_U_spectral}) $\mu$ is sufficiently
large. Put $V=1+V_{1}$ where $V_{1}$ has sufficiently small $l_{\infty}$-norm,
then 
\[
a(\sqrt{V})=\frac{c}{\mu^{2}+V}=\frac{c}{\mu^{2}}(1-\frac{1}{\mu^{2}}V_{1}+o(\frac{1}{\mu^{2}}))
\]
and the linear in $V_{1}$ term provides non-zero correlations $<p_{i}p_{j}>,i\neq j$,
if $V_{1}(i,j)\neq0$. 
\item As a rare exception, one can construct, using (\ref{Gibbs-nonGibbs}),
even for $N=1,2$, examples of $H\in\mathbf{H}_{G}$ and $C_{f}\in S$
with Gibbs limiting distribution. We do not know whether such kind
of examples have physical sense. 
\item We did not consider here other generalized processes with independent
values - derivatives of the white noise and of the (non-gaussian)
Levy processes. It is an open question what limiting distribution
will be for these <<no-memory>> cases. It seems that the white noise
is the only stationary gaussian process, providing convergence of
the system to Gibbs states for almost any $V$. \end{enumerate}

\end{document}